\documentclass[a4,11pt]{article}
\pdfoutput=1 

\usepackage{jcappub}
\usepackage[utf8]{inputenc}
\usepackage{graphicx}
\usepackage{color, graphics}
\usepackage{amsmath,amssymb,amsfonts}
\usepackage{verbatim}   
\usepackage{subfigure}  
\usepackage{acronym}
\usepackage{enumerate}   
\usepackage{hyperref}
\usepackage{mathrsfs}
\usepackage{multirow}
 \usepackage{cancel}
 \usepackage{url}
  \usepackage{slashed,afterpage}
 \usepackage{hyperref}
 \usepackage{units}
\usepackage{enumerate}  
\emailAdd{pchand31@uic.edu} 
\emailAdd{unwin@uic.edu}

\def\beq{\begin{equation}\begin{aligned}}
\def\eeq{\end{aligned}\end{equation}}

\begin{document}

\title{\huge Decoupling of Asymmetric Dark Matter During an Early Matter Dominated Era}
\author{Prolay Chanda}
\author{and James Unwin}
\affiliation{Department of Physics,  University of Illinois at Chicago, Chicago, IL 60607, USA}

\abstract{In models of Asymmetric Dark Matter (ADM) the relic density is set by a particle asymmetry in an analogous manner to the baryons. Here we explore the scenario in which ADM decouples from the Standard Model thermal bath during an early period of matter domination. We first present a model independent analysis for a generic ADM candidate with  s-wave annihilation cross section with fairly general assumptions regarding the origin of the early matter dominated period. We contrast our results to those from conventional ADM models which assume radiation domination during decoupling. Subsequently, we examine an explicit example of this scenario in the context of an elegant SO(10) implementation of ADM in which the matter dominated era is due to a long lived heavy right-handed neutrino.  In the concluding remarks we discuss the prospects for superheavy ADM in this setting.}

\maketitle


\vspace{5mm}
\section{Introduction}

Asymmetric Dark Matter (ADM) \cite{Nussinov:1985xr,Gelmini:1986zz,Chivukula:1989qb} draws a direct analogy with the mechanism through which the present-day density of baryons is set. The dark matter carries a conserved quantum number (analogous to baryon number $B$), and there is an asymmetry in the number of dark matter particles $\chi$ and its antiparticle $\overline{\chi}$.  The asymmetry is typically defined as follows
\beq\label{eq:eta}
\eta_{\chi}\equiv \frac{n_{\chi}-n_{\overline{\chi}}}{s},
\eeq
where $n$ indicates the number density,  and $s$ is the Standard Model entropy density. Provided the dark matter annihilation rate is sufficiently large, then the relic density is not determined by the point of decoupling (as in freeze-out dark matter \cite{Zel'dovich,Chiu}), but rather is set by the size of the asymmetry, such that $Y_{\chi}\simeq\eta_{\chi}$ and $Y_{\overline{\chi}}\ll Y_{\chi}$ (for reviews of ADM see \cite{Zurek:2013wia,Petraki:2013wwa}).

\begin{figure}[t]
    \begin{center}
 \includegraphics[width=0.75\textwidth]{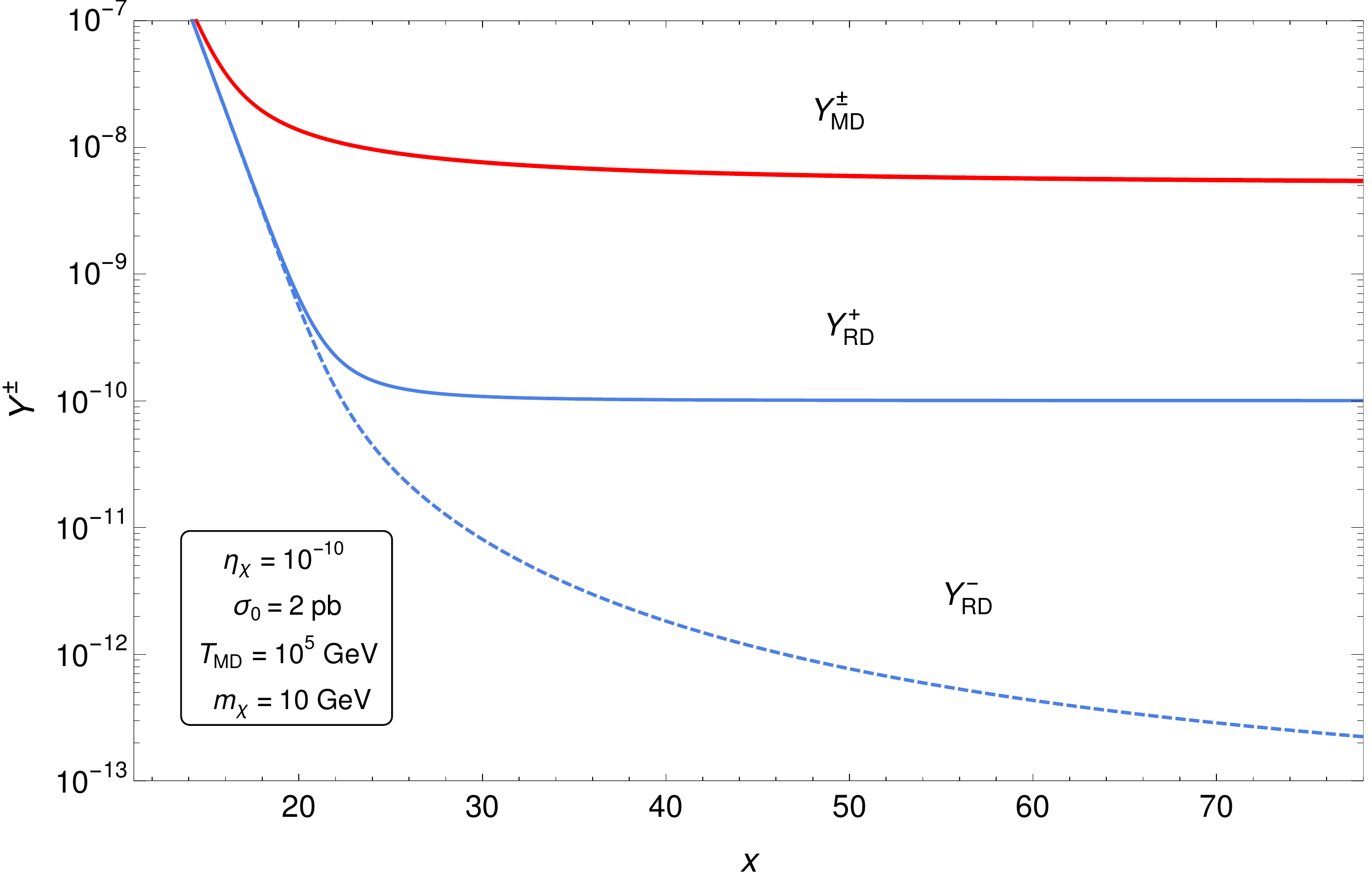}
    \end{center}
    \vspace{-3mm}
     \caption{Evolution of the abundances of dark matter  $Y^{+}$  and anti-dark matter $Y^{-}$  as functions of $x = m_{\chi}/T$, assuming radiation domination (RD) and matter dominated (MD), for an annihilation cross-section $\sigma_{0} = 2~{\rm pb}$ and an initial dark matter asymmetry $\eta_{\chi}= 10^{-10}$. We assume that matter domination occurs at a temperature of $10^{5}~{\rm GeV}$ in the case of decoupling during a matter dominated era. Observe that for the same particular choices of particle physics parameters (masses, couplings),  radiation dominated decoupling implies an ADM scenario with the late time abundance set by $\eta_{\chi}$, while the  matter dominated case leads to symmetric dark matter. Thus cosmology can play a role in determining the late time dark matter scenario. }
 \label{fig:abundance2}
 \end{figure}

For dark matter that decouples in a matter dominated era, rather than during radiation domination, there are two main consequences. First, the Hubble rate is different, and this alters the dynamics of decoupling \cite{Hamdan:2017psw}. Secondly, entropy production can occur when radiation domination is restored, which should occur prior to Big Bang Nucleosynthesis (BBN), this either dilutes the dark matter abundance or leads to dark matter production. Indeed, one can see the difference in decoupling between the two scenarios by inspection of Figure~\ref{fig:abundance2}. This figure shows the evolution of the dark matter abundance $Y_{\chi}\equiv n_{\chi}/s$  and the anti-dark matter abundance $Y_{\bar{\chi}}$ as a function of $x=m_{\chi}/T$ where $T$ is the temperature. This is shown for both the case of freeze-out during radiation domination $\left(Y_{\chi,\bar{\chi}}\right)_{\rm RD}$ and freeze-out assuming matter domination $\left(Y_{\chi,\bar{\chi}}\right)_{\rm MD}$. Notably, for the same parameter values the relic density is asymmetric with $\left(Y_{\bar{\chi}}\right)_{\rm RD}\ll \left(Y_{\chi}\right)_{\rm RD}$ for radiation dominated freeze-out, while for matter dominated freeze-out the final dark matter abundance consists of symmetric contributions from the dark matter and anti-dark matter: $\left(Y_{\bar{\chi}}\right)_{\rm MD}\simeq \left(Y_{\chi}\right)_{\rm MD}$.

This paper is structured as follows; we first adopt a model independent perspective and obtain a semi-analytic solution to the Boltzmann equation for the case of ADM with an s-wave annihilation cross-section which decouples during an early period in which the universe is matter dominated. In particular, we calculate the abundances of the dark matter and anti-dark matter at the point of decoupling.  Then, in Section \ref{Sec4}, we discuss the importance of the entropy injection on the relic density of ADM and thus identify regions of parameter space for which this scenario successfully reproduces the observed dark matter relic abundance. Following this, in Section \ref{Sec2} we present an explicit implementation involving an ADM candidate which elegantly arises from an  SO(10) unified theory and in which the matter dominated era is due to a long-lived heavy right-handed neutrino. We numerically calculate the relic density of ADM in this scenario and show that the assumption of dark matter decoupling during matter domination makes the scenario significantly more attractive. In Section \ref{Sec5}, we give some concluding remarks and discuss the prospect of realizing  Superheavy ADM in the case that decoupling occurs during a matter dominated era.

\section{Boltzmann Analysis Assuming Matter Domination} 
\label{Sec3}

We first consider a more model independent approach in which the early matter dominated era is sourced by a general population of matter-like states and where we remain agnostic regarding the manner of dark matter freeze-out apart from the assumption that it is via  s-wave annihilations. We start by outlining the differences between radiation dominated freeze-out and matter dominated freeze-out. We then quantify the impact on models of ADM by deriving the freeze-out abundance assuming that decoupling occurs during a matter dominated era. In Section \ref{Sec4} we will discuss the entropy dilution of the freeze-out abundance due to the transition to radiation domination, leading to the final relic density of dark matter.
 
 \newpage
  \subsection{Matter Dominated Cosmology}
  
  For dark matter decoupling during a matter dominated era, the Hubble rate at freeze-out has a different temperature dependence compared to during a radiation dominated era \cite{Hamdan:2017psw}.  We consider the presence of a decoupled scalar field\footnote{To be explicit the field $\phi$ should not be identified as the inflaton.} $\phi$ along with the visible sector and dark matter $\chi$.  Just after inflationary reheating, $\chi$ and $\phi$ behave like radiation until a characteristic temperature $T_{\star}$ when $\phi$ starts evolving matter-like. A long-lived $\phi$  results in a matter dominated early universe. If $T_{\star}\gg T_{f}\gg T_\Gamma$, the dark matter freezes out in a matter dominated universe. The Hubble parameter for $T_{\star}>T>m_{\chi}$ can be defined as
  \beq\label{eq:H1}
H^{2}&
= H^{2}_{\star}\left[
\frac{r~g_{\ast}}{g_{\ast}+g_{\chi}}\left(\frac{a_{\star}}{a}\right)^{4}+(1-r)\left(\frac{a_{\star}}{a}\right)^{3}+ \frac{r~g_{\chi}}{g_{\ast}+g_{\chi}}\left(\frac{a_{\star}}{a}\right)^{4} \right],
\eeq
where the subscript `$\star$' refers to the value of a physical quantity evaluated at $T = T_{\star}$,  $H_{\star} = \sqrt{4\pi^{3}/45}~k^{1/2}g_*^{1/2}(T^{2}_{\star}/M_{\rm Pl})$, and $g_{\ast}$ is the effective number of relativistic degrees of freedom of the Standard Model and $g_{\chi}$ is the internal degrees of freedom of the DM, the factor $k = 1$ for bosonic   dark matter and $k=7/8$ for fermionic dark matter. The quantity $r$ accounts for the fraction of the energy density in radiation at $a=a_*$. Note also that provided the thermal bath entropy is conserved, the scale factor $a(T)$ can be related to the bath temperature $T$ as follows 
\beq\label{eq:a2T}
 \frac{a(T_{\star})}{a(T)}\simeq \left(\frac{g_{\ast}(T)}{g_{\ast}(T_{\star})}\right)^{1/3}\frac{T}{T_{\star}}.
 \eeq
From eq.~\eqref{eq:H1} and using eq.~\eqref{eq:a2T}, it can be seen that for $r\simeq 1$, the Hubble parameter $H\propto T^{2}$, and therefore the universe is radiation dominated.  On the other hand, for $r\ll 1$, the Hubble parameter $H\propto T^{3/2}$, corresponding to a matter dominated era at $a=a_\star$. If the universe is radiation dominated while the $\phi$ field becomes non-relativistic at temperature $T_{\star}$, the $\phi$ field contribution to the energy density of the universe grows with time potentially leading to a matter dominated universe. We denote the temperature of the thermal bath at which the universe becomes matter dominated as $T_{\rm MD}$ (specifically, the point at which the $\phi$ field accounts for half of the total energy density of the universe), this can be related to $T_\star$ via the following relationship
\beq \label{MDD}
T_{\rm MD} = \frac{(1-r)}{r}\left(\frac{g_{\ast}(T_{\star})}{g_{\ast}(T_{\rm MD})}\right)^{1/4}T_{\star}.
\eeq

For both  matter dominated and the radiation dominated cosmology,   $H(t)$ scales as $({\rm time})^{-1}$. This motivates us to adopt the following form for $H$ 
\beq\label{eq:H3}
H = \frac{\nu}{t},
\eeq
with   $\nu = 2/3$ and 1/2 for matter domination and radiation domination, respectively.
Note that we will assume here that dark matter decoupling occurs well before the subsequent transition from matter domination to radiation domination due to $\phi$ decays. The converse case leads to significant differences in the Boltzmann analysis, see \cite{Giudice:2000ex}. We discuss for what cases this assumption holds in Appendix \ref{APD} (see \cite{Chanda:2019xyl} for a more detailed discussion).

\subsection{The Asymmetric Yield}

It is helpful to define   the fractional asymmetry $F = Y_{\chi^{\dagger}}/Y_{\chi}$,  being the abundance of anti-dark matter compared to the dark matter   \cite{Graesser:2011wi} . After dark matter decouples, at the freeze-out temperature $T_f$, the fractional asymmetry is constant $F_{f}\equiv F(T_f)$ (in the absence of any subsequent production of the dark matter). Thus $F=1$ implies conventional symmetric dark matter, while $F<1$ corresponds to ADM. The relic abundance could be mildly asymmetric, e.g.,~$F\sim0.1$, in which case the abundance of the anti-dark matter is depleted by an order of magnitude only, or it could be highly asymmetric $F\ll1$ similar to the baryons. Notably, $F=F(\sigma, m_\chi)$ and a smaller fractional asymmetry requires a larger annihilation cross-section $\sigma$, potentially implying better detection prospects, but otherwise, $F$ is a free parameter. 

We next derive an expression for the evolution of the fractional asymmetry $F$, thus extending a result of  \cite{Graesser:2011wi}  to the case of ADM which decouples during a period of matter domination.
Writing $Y_{\chi}\equiv Y^{+}$ and $Y_{\chi^{\dagger}}\equiv Y^{-},$ the Boltzmann equation for dark matter and anti-dark matter can be expressed as follows \cite{Iminniyaz:2011yp} (see also \cite{MarchRussell:2012hi,Iminniyaz:2018das,Griest:1986yu}) 
 \beq\label{be1}
 \frac{dY^{\pm}}{dt} = -\langle\sigma v\rangle s(T)(Y^{+}Y^{-}-Y^{+}_{ {\rm eq}}Y^{-}_{ {\rm eq}}),
 \eeq
where  $\langle\sigma v\rangle$ is dark matter self annihilation cross-section and  $s$ is the entropy density 
$ s(T) = \frac{2\pi^{2}}{45}g_{\ast,s}(T)T^{3}$ in terms of
   $g_{\ast,s}(T)\simeq g_{\ast}(T)$  the effective number of relativistic degrees of freedom.
Finally, the equilibrium abundances can be expressed as follows \beq\label{eq:F3}
Y_{\rm eq}^{\pm}&=bx^{3/2}e^{-x}e^{\pm\mu/T},
\eeq
where $ b = \frac{45g}{4\sqrt{2}\pi^{7/2}g_{\ast,s}} $ and $g$ is the dark matter internal degrees of freedom (e.g.~$g=2$ for complex scalar or $g=4$ for Dirac fermion). Here $\mu$ is the $\chi$ chemical potential which characterizes the difference between dark matter and anti-dark matter, such that in the nonrelativistic limit $\eta_\chi\propto \sinh(\mu_\chi/T)$.

   The dark matter annihilation cross-section can be expanded in a power series around $x=0$, $\langle\sigma v\rangle=\sum_{n=0}^{\infty}\sigma_{n}x^{-n}$, expanding to second order (i.e.~s-wave $\sigma_0$ and p-wave $\sigma_1$ terms) one has
 $\langle\sigma v\rangle=\sigma_{0}+\sigma_{1}x^{-1}$.
Then from eqns.~\eqref{eq:H1} and \eqref{eq:H3}  we can rewrite \eqref{be1} as follows
\beq\label{eq:be3}
 \frac{dY^{\pm}}{dx}& =-\sum_{n}\lambda_{n} x^{-n-2}g_{\rm eff}^{1/2}\left(Y^{+}Y^{-}-Y^{+}_{\rm eq}Y^{-}_{\rm eq}\right),
\eeq
 where the quantity $\lambda_{n}$ contains the cross section and
   is  defined as
\beq\label{lambda}
\lambda_{n} &
= \nu k^{-1/2}\sqrt{\frac{4\pi}{45}}m_{\chi}M_{\rm Pl}\sigma_{n}.
\eeq

We obtain numerical solutions to eq.~\eqref{eq:be3} for a given dark matter annihilation cross-section $\sigma_{0}$ (neglecting the higher-order contributions) and fixing the dark matter mass $m_{\chi}$, the critical temperature $T_{\star}$ and the initial dark matter asymmetry $\eta_{\chi}$. An example evolution was given in Figure \ref{fig:abundance2}. Observe that in the scenario of dark matter decoupling  during  radiation domination, the freeze-out dark matter abundance is determined by the initial asymmetry $\eta_{\chi}$ for this particular choice of  parameters. In contrast,  in the matter dominated freeze-out case, with otherwise the same parameter values, the dark matter abundance at decoupling is comparable to the anti-dark matter abundance thus implying a symmetric dark matter scenario.

Furthermore, it can be insightful to recast the evolution of the cosmic abundances of the dark matter in terms of the fractional asymmetry $F = Y^{-}/Y^{+}$. Specifically, from eq.~\eqref{eq:eta}, the equilibrium abundance eq.~\eqref{eq:F3}  follows from the geometric mean of the equilibrium abundance of the dark matter and anti dark matter in terms of $F_{{\rm eq}} \equiv \frac{Y^{-}_{{\rm eq}}}{Y^{+}_{{\rm eq}}}$ as
\beq\label{F2}
Y_{\rm eq}\equiv\sqrt{Y_{\rm eq}^{-}Y_{\rm eq}^{+}} = \sqrt{\frac{F_{\rm eq}\eta_{\chi}^{2}}{(1-F_{\rm eq})^{2}}}.
\eeq
This implies that the equilibrium fractional asymmetry evolves according to 
\beq\label{F1}
& F_{\rm eq} = \exp\left(-2\sinh^{-1}\left(\frac{\eta_{\chi}}{2Y_{\rm eq}}\right)\right).
\eeq
Moreover, the evolution of dark matter and anti-dark matter $Y^{\pm}$  can be expressed as follows
 \beq\label{eq:be0}
 \frac{dY^{\pm}}{dx}&=-\langle\sigma v\rangle \frac{dt}{dx}s(T)\left(\frac{\eta_{\chi}^{2}F}{(1-F)^{2}}-Y^{+}_{\rm eq}Y^{-}_{\rm eq}\right)~.
 \eeq
 Thus in the early universe, much before dark matter decoupling occurs, it is expected that  $F \simeq F_{\rm eq}$. On the other hand, at later times after decoupling (i.e.~$x\gg  x_{f}$), the RHS of the eq.~\eqref{eq:be0} will be dominated by the first term since $Y_{\rm eq}$ is exponentially dependent on $x$. 
From eq.~(\ref{eq:be3}) it follows that one can express the evolution of the fractional asymmetry as 
\beq\label{c}
 \frac{dF}{dx}
  &=- \eta_{\chi} g_{\rm eff}^{1/2}\sum_{n}\lambda_{n}x^{-n-2}\left(F-F_{\rm eq}\left(\frac{1-F}{1-F_{\rm eq}}\right)^{2} \right),
  \eeq
    where $g_{\rm eff}$ is the effective number of relativistic degrees of freedom given by
\beq\label{gcases}
g_{\rm eff}^{1/2}\simeq g_{\ast}^{1/2}(T_{\star})\left(\frac{3}{4}\frac{T_{\star}(1-r)x}{m_{\chi}}+r\right)\left(\frac{T_{\star}(1-r)x}{m_{\chi}}+r\right)^{-3/2}~.
\eeq  
 Since we are interested in times much before BBN, we take $g_*$, to be constant and set it to be Standard Model value at temperatures above the top mass $g_{\ast,s}(x)\simeq g_{\ast}(x)\simeq g_{\ast}(x_{\star})\simeq 106.75$. Note also that for the case of radiation dominated freeze-out one has $g_{\rm eff}^{1/2}\simeq g_*^{1/2}$.

\vspace{-2mm}
\subsection{Dark Matter Decoupling}
\vspace{-1mm}
\label{3.3}

 At the point of dark matter decoupling, the two terms on the RHS of eq.~\eqref{eq:be0} cease to balance, however to good approximation $\frac{{\rm d}F}{{\rm d}x}\approx \frac{{\rm d}F_{\rm eq}}{{\rm d}x}$, and it follows that
\beq\label{eq:F4}
F^{\prime}_{\rm eq} \approx -\sum_{n}\delta_{n}\lambda_{n}\eta_{\chi} g_{\rm eff}^{1/2}x^{-n-2}F_{\rm eq},
\eeq
  where the prime indicates a derivative with respect to the $x$ and  $\delta_{n}$  is a constant fixed by fitting the numerical solution, normally taken to be $(n+1)$ \cite{Scherrer:1985zt}.
  
Next, we derive the temperature of dark matter decoupling $x_f$. We use eq.~\eqref{F2}, along with eq.~\eqref{eq:F4},  to re-express the condition of eq.~\eqref{F1} in the following manner
\beq\label{eq:F5}
\left(1-\frac{3}{2x_{f}}\right)\frac{1-F_{\rm eq,f}}{1+F_{\rm eq,f}}\approx - \eta_{\chi} g_{\rm eff}^{1/2}\sum_{n}\delta_{n}\frac{\lambda_{n}}{x_{f}^{n+2}}.
\eeq
Restricting to the relevant limit in which  $\eta_{\chi} g_{\rm eff}^{1/2}\sum\limits_{n}\lambda_{n}x_{f}^{-n-2}\ll 1$, we can obtain an iterative approximate solution for the point of decoupling
\beq\label{xf0}
x_{f}= &\ln(\delta_{n} g_{{\rm eff},f}^{1/2}b\lambda_{n})+\frac{1}{2}\ln\left(\frac{\ln^{3}(\delta_{n} g_{{\rm eff},f}^{1/2}b\lambda_{n})}{\left(\ln^{2n+4}(\delta_{n} g_{{\rm eff},f}^{1/2}b\lambda_{n})-\frac{1}{4}\left(\delta_{n}\lambda_{n}\eta_{\chi}\right)^{2}g_{{\rm eff},f}\right)}\right).
\eeq
 Using eqns.~\eqref{gcases} and \eqref{lambda} with $r \leq 0.99$ (which implies that there is a non-zero contribution from the matter-like species $\phi$ to the energy density at $T=T_\star$) we obtain the following expression 
\beq\label{M}
   x_{f}=\ln \mathcal{M}
+\frac{1}{2}\ln\left(\frac{\ln^{3}\mathcal{M}}{\ln^{2n+4}\mathcal{M}-\mathcal{N} } \right)
\eeq
in terms of
 \beq\label{eq:xfmd}
& \mathcal{M}=\frac{\sqrt{45}g}{\sqrt{32}\pi^{3}}(n+1)k^{-1/2}g_*^{-1/2}(1-r)^{-1/2}M_{\rm Pl}m_{\chi}^{3/2}T_{\star}^{-1/2}x_{f}^{-1/2}\sigma_{n} \\
& \mathcal{N}=\left(\sqrt{\frac{\pi}{180}}(n+1)k^{-1/2}(1-r)^{-1/2}\eta_{\chi} M_{Pl}m_{\chi}^{3/2} T_{\star}^{-1/2}x_{f}^{-1/2}\sigma_{n}\right)^{2} g_{\ast}(T_{\star}),
 \eeq
  where the dependence of $\mathcal{M}$ on $x_{f}$ comes from the fact that $g_{{\rm eff}}$ for the matter dominated case is a function of $x_{f}$,
   cf.~eq.~(\ref{gcases}). Recall, $k$ distinguishes between the bosonic and fermionic dark matter,
    and $r$ is the fraction of the energy density contained in radiation at $T=T_\star$.

\section{The Transition to Radiation Domination and the ADM Relic Abundance }
\label{Sec4}

In the scenario in which the dark matter freezes out during matter domination in the presence of the decoupled heavy scalar field $\phi$,  then $\phi$ must subsequently decay in order to restore radiation domination prior to BBN to reproduce cosmological observables. The decay of $\phi$ dilutes\footnote{In principle, $\phi$ decays could produce dark matter and anti-dark matter. In particular, this could occur via loops of Standard Model particles, as discussed in  \cite{Chanda:2019xyl,Kaneta:2019zgw}. While one might be concerned that even if the production of dark matter from decays is negligible, anti-dark matter production might alter the fractional asymmetry $F$,  in Appendix \ref{DMprodLoop} we argue that this is not the case in the parameter regions of interest.} the freeze-out dark matter abundance by a factor $\zeta$ and, furthermore, the reheat temperature $T_{\rm RH}$ is linked to the dilution factor $\zeta$. Specifically, under the assumption that $\phi$ decays suddenly when the thermal bath has a temperature $T_\Gamma$, such that the reheating temperature of the Standard Model thermal bath following $\phi$ decays is $T_{\rm RH} \sim\sqrt{\Gamma M_{\rm Pl}}$, then the dilution factor $\zeta$ can be expressed as 
\beq\label{eq:DilutionFactor0}
\zeta = \frac{s_{\rm before}}{s_{\rm after}}\simeq\left(\frac{T_\Gamma}{T_{\rm RH}}\right)^{3}.
\eeq
 This is interesting because the entropy injection from the $\phi$ decays allows the dark matter to be overabundant at freeze-out and subsequently diluted in order to obtain the observed dark matter relic density. For $\zeta\ll 1$, smaller dark matter annihilation cross-sections or larger dark matter masses are allowed, therefore weakening search constraints and also unitarity limits  \cite{Griest:1989wd} (typically $m_{\chi}\lesssim 100$ TeV).

We can derive the evolution of the scale factor at critical temperature $a_{\star}$ to the point $a_{\Gamma}$  (at which the $\phi$ states decay simultaneously), using $H\simeq \Gamma$ in eq.~\eqref{eq:H1} we obtain
    \beq
    \left(\frac{a_{\star}}{a_{\Gamma}}\right)^{3}\approx \left(\frac{\Gamma}{H_{\star}(1-r)^{1/2}}\right)^{2/3}.
    \eeq
    Taking this with $T_{\rm RH} \sim\sqrt{\Gamma M_{\rm Pl}}$, we can express  $T_{\Gamma}$ in terms of $T_{\rm RH}$ and $T_{\star}$ 
    \beq
    T_{\Gamma}\simeq \left(\frac{45}{4\pi^{3}g_*(1-r)}\frac{T_{\rm RH}^{4}}{T_{\star}}\right)^{1/3}~.\label{111}
    \eeq
  Moreover, using eq.~(\ref{eq:DilutionFactor0}) we can obtain a formula which relates $\zeta$ to $T_{\rm RH}$ and $T_{\star}$ 
    \beq
\zeta  \simeq \frac{45}{4\pi^{3}}   \frac{T_{\rm RH} }{(1-r)T_{\star}g_*} \sim  \frac{T_{\rm RH} }{(1-r) g_\star M_\phi} ~,\label{000}
    \eeq
        where in the latter expression we have taken that $\phi$ becomes non-relativistic at $T_{\star}\sim M_\phi$ and we have dropped the $\mathcal{O}(1)$ numerical prefactor. 
        
\subsection{The Dark Matter Relic Abundance}
\label{sec:2}

In Section \ref{3.3} we derived an expression for the point of dark matter decoupling, eq.~(\ref{M}), we now use this result to derive the dark matter relic abundance. We start from an approximated analytic solution to eq.~\eqref{eq:be3} (which we derive in Appendix \ref{AnalyticYX}), specifically, for $x\gg x_{f}$ the anti-dark matter abundance is given by
 \beq\label{eq:ADMabundance0}
 Y^{-}_{f} &\simeq\frac{\eta_{\chi}}{\exp\left(\eta_{\chi}\sum\limits_{n}\lambda_{n}\Phi_{n}(\infty , m_{\chi})\right)-1},
 \eeq
 where $\Phi_{n}(\infty , m_{\chi})$ is defined as
 \beq \label{Phi}
\Phi_{n}(x,m_{\chi}) &= \int_{x_{f}}^{x}dx'x'^{-n-2}g_{\rm eff}^{1/2}.
\eeq
It then follows that the total dark matter density at decoupling can be expressed as follows
\beq\label{eq:omega1}
\Omega_{\rm DM,f}h^{2} = \frac{s_{0}m_{\chi}}{\rho_{c}}h^{2}\left(Y_{f}^{+}+Y^{-}_{f}\right),
\eeq
where $\rho_{c} = 3H_{0}^{2}M_{\rm Pl}^{2}/(8\pi)$ 
with $M_{\rm Pl}=1.221\times 10^{19} ~{\rm GeV}$ and $s_{0}\approx 2970 ~{\rm cm}^{-3}$.

 Provided dark matter production after the freeze-out is negligible, the dark matter relic abundance corresponds to the abundance at decoupling diluted by a factor $\zeta$ due to the  entropy injection of $\phi$ decays. The $\zeta$ factor is given in eq.~\eqref{eq:DilutionFactor0} and this lead to the following expression for the relic density of the dark matter 
 \beq\label{eq:omega11}
\Omega_{\rm FO}h^{2} = \zeta\frac{s_{0}m_{\chi}}{\rho_{c}}h^{2}\left(2Y^{-}_{f}+\eta_{\chi}\right).
\eeq
 Therefore, using eq.~\eqref{eq:ADMabundance0} the dark matter relic density after decoupling during a matter dominated era and subsequent entropy dilution is given by
\beq\label{eq:omega4}
\Omega^{\text{Relic}}_{\rm  DM}h^{2}&\simeq  
\frac{5.7\times 10^{8}\zeta \eta_{\chi}m_{\chi}~{\rm GeV}^{-1}}{\exp\left(\sqrt{\frac{\pi}{45}}k^{-1/2}g_*^{1/2}\frac{\eta_{\chi}}{\sqrt{1-r}}m_{\chi} M_{\rm Pl}\sum\limits_{n}\frac{\sigma_{n}x_{f}^{-(n+\frac{3}{2})}}{(n+\frac{3}{2})}x_{\star}^{1/2}\right)-1}
+2.85\left(\frac{m_{\chi}}{\text{\rm GeV}}\right)\zeta\left(\frac{\eta_{\chi}}{10^{-8}}\right) ~.
\eeq
For  $\eta_{\chi}\ll (\sum\limits_{n}\lambda_{n}\Phi_{n}(\infty,m_{\chi}))^{-1}$   the expression above can be simplified to obtain
\beq\label{eq:omega5}
\Omega^{\text{Relic}}_{\rm DM}h^{2}&\simeq 
 \zeta\left[2.16\times 10^{9}\frac{k^{1/2}}{g_*^{1/2} M_{\rm Pl}}\left(\sum\limits_{n}\frac{\left(1-r\right)^{-1/2} \sigma_{n}{{\rm GeV}}}{(n+3/2)x_{f}^{n+3/2}x_{\star}^{-1/2}}\right)^{-1}\right]+2.85\left(\frac{m_{\chi}}{\text{\rm GeV}}\right)\zeta\left(\frac{\eta_{\chi}}{10^{-8}}\right).
\eeq
Note that the first term on the RHS corresponds to the symmetric contribution to the relic density and the second term represents the asymmetric component.

The dark matter relic density today is set by both the symmetric and the asymmetric contributions. For a certain choice of parameters the symmetric part annihilates away, and the dark matter density today is set by the asymmetric part. The final asymmetry, as we observe today, is the initial asymmetry followed by the entropy injection
\beq
\eta_{\chi}^{\rm final} = \zeta \eta_{\chi}^{\rm initial}.
\eeq
Notably, the entropy injection also dilutes the initial baryon asymmetry $\eta_{B}^{\rm initial}$  to the final baryon asymmetry $\eta_{B}^{\rm final}$ in a similar fashion. If we suppose that the initial asymmetry can be no larger than $\eta_{B}^{\rm initial}\sim\mathcal{O}(1)$ (as argued in \cite{Linde:1985gh}) then the observed baryon asymmetry today $\eta_{B}\sim 10^{-10}$ implies a weak bound on the magnitude of $\zeta$.

\subsection{The Fractional Dark Matter Asymmetry}
From eq.~\eqref{eq:omega11}, the relic density of dark matter is the decoupling abundance multiplied by the dilution factor $\zeta$
which can be  expressed in terms of $F_{f}$ as follows
\beq\label{eq:omega3}
\Omega^{\rm Relic}_{\rm DM}h^{2} =\zeta \frac{s_{0}m_{\chi}}{\rho_{c}}h^{2}\left(\eta_{\chi}+\frac{2\eta_{\chi}F_{f}}{1-F_{f}}\right),
\eeq
where the first term corresponds to the asymmetric part. To proceed, we re-express the evolution of the fractional asymmetry, as given in eq.~\eqref{eq:F4}, in terms of $\Phi$ and $ F_{\rm eq}$ as follows
\beq\label{eq:Fdef}
F=F_{\rm eq,f}~\exp\left(-\sum\limits_{n}\lambda_{n}\eta_{\chi}\Phi_{n}(x,m_{\chi})\right).
\eeq
In the case of symmetric dark matter ($F_{f} \to 1$) the relic density however should depend strongly on the cross-section, which can be seen by the fact that  fractional asymmetry eq.~\eqref{eq:Fdef}   depends strongly on the cross-section $\sigma_{n}$ and in the symmetric dark matter limit the fractional asymmetry can be approximated as 
\beq
F_{f}\approx 1-\sum\limits_{n}\lambda_{n}\eta_{\chi}\Phi_{n}(\infty ,m_{\chi} )~,
\eeq
 where the parameter $\lambda_{n}$ is directly proportional to the cross-section $\sigma_{n}$. 
Moreover, in the case where the initial asymmetry is sufficiently small such that $Y^{+}_{f}\simeq Y^{-}_{f}\gg \eta_{\chi}$,
and therefore, $F_{\rm eq,f}\approx 1$, then the fractional asymmetry simplifies to 
\beq\label{eq:Finf}
F_{f} \simeq \exp\left(-\sum\limits_{n}\lambda_{n}\eta_{\chi}\Phi_{n}(\infty ,m_{\chi} )\right).
\eeq
However, as suggested in \cite{Graesser:2011wi}, the above form remains a valid approximation for stronger initial asymmetries  provided $F_{f}\gtrsim 10^{-4}$. Note that working with $F$ and $\eta_\chi$ is equivalent to working with $Y_+$ and $Y_-$, in either case the physics is controlled by two parameters.

Furthermore,  the  relic density of baryons today can be written as follows
\beq
\Omega^{\rm Relic}_{B} h^{2}= \zeta \frac{s_{0}m_{p}}{\rho_{c}}h^{2}\eta_{B}, 
\eeq
where $m_{p}$ is the proton mass and $\eta_{B}$ is the initial baryon asymmetry.
It follows that the ratio of the dark matter to the baryon relic density can be expressed in the form \cite{Graesser:2011wi}
\beq\label{eq:ObsOmega1}
\frac{\Omega^{\rm Relic}_{\rm DM}}{\Omega^{\rm Relic}_{B}} = \left(\frac{1+F_{f}}{1-F_{f}}\right)\frac{m_{\chi}\eta_{\chi}}{m_{p}\eta_{B}}.
\eeq

\subsection{Dark Matter with s-wave Annihilations}

\begin{figure}[t!]
\centerline{\includegraphics[width=0.6\textwidth]{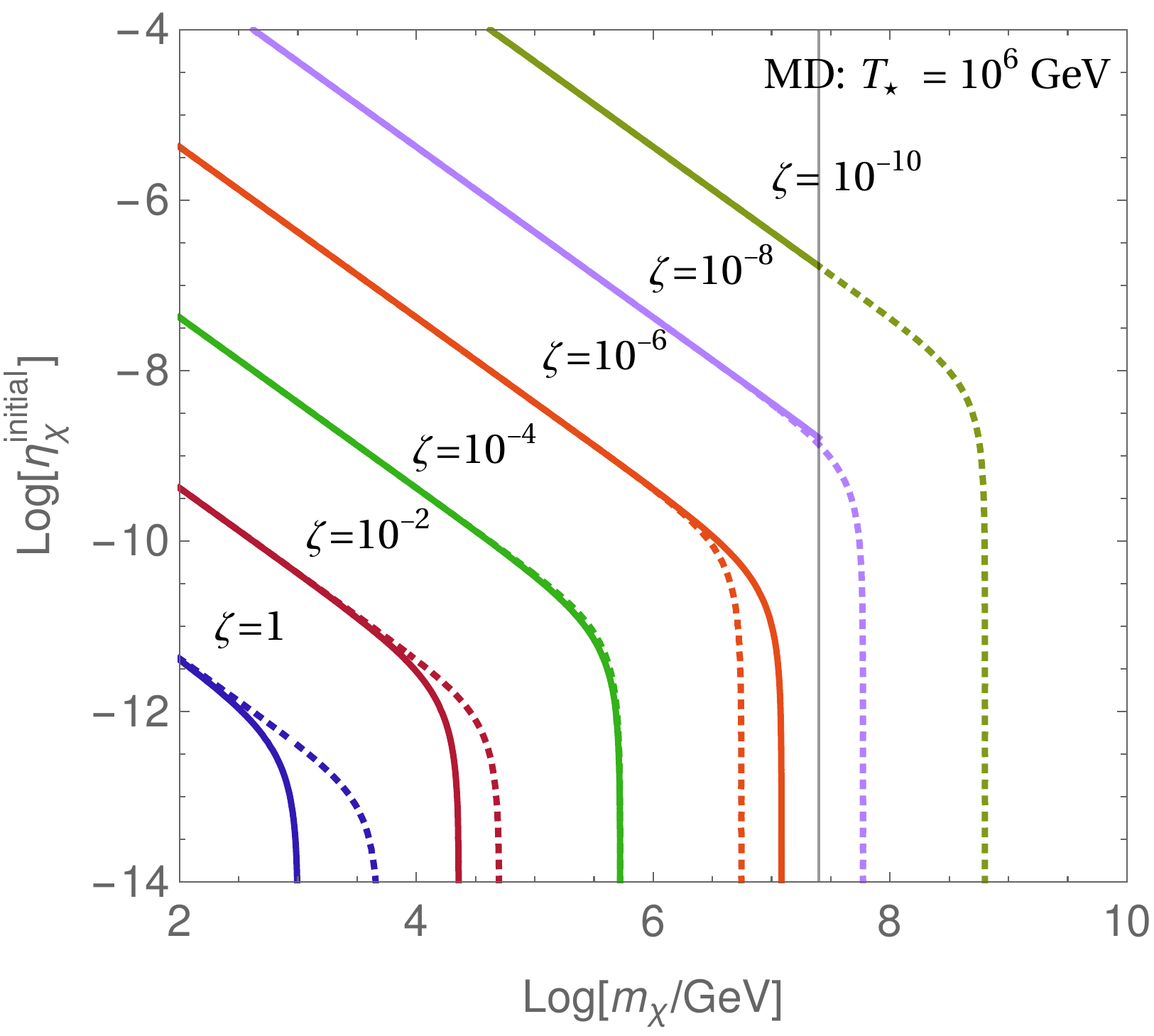}}
\caption{Contours for which the observed relic density $\Omega_{\rm relic}h^{2}\approx 0.12$ is obtained as functions of the initial asymmetry $\eta_{\chi}^{\rm initial}$ and the mass of dark matter $m_{\chi}$ for different entropy dilution factor $\zeta$ as indicated.  With $\sigma_{0}$ as in  eq.~\eqref{eq:MIsigma}, the plot shows the case of two different values of the coupling constant $\kappa_{\chi}=0.3$ and 0.05 as indicated by solid and dashed lines, respectively.  We cut off matter domination case for $m_{\chi}>25\times T_\star$ since $m_{\chi}/25$ is the characteristic freeze-out point and thus beyond this the dark matter freezes  out in a radiation domination regime prior to $\phi$ matter domination.}
\label{fig:omega1}
\end{figure}

We will consider the simplest case of dark matter with an s-wave annihilation channel to Standard Model states which we parameterise in a model independent manner as follows
\beq\label{eq:MIsigma}
\sigma_{0}\approx \frac{\kappa^{2}}{m_{\chi}^{2}}.
\eeq
Thus the dark matter relic density eq.~\eqref{eq:omega5} is a function of the initial dark matter asymmetry $\eta^{\rm initial}_{\chi}$, the entropy dilution $\zeta$, the dark matter mass $m_{\chi}$, the single coupling constant $\kappa$ and in the matter dominated scenario, the critical temperature $T_{\star}$. We illustrate this in Figure~\ref{fig:omega1} which presents contours of the observed dark matter relic density as a function of the initial asymmetry $\eta_{\chi}^{\rm initial}$ and the dark matter mass $m_{\chi}$ for different entropy dilution factors $\zeta$, for two choices of the coupling  $\kappa_{\chi}=0.3$ and 0.05 and two values for the critical temperature $T_\star$.
For the plots of this section we assume that the dark matter is a scalar boson (the difference for fermion dark matter is minimal).

\begin{figure}[t!]
\centerline{\includegraphics[width=0.5\textwidth]{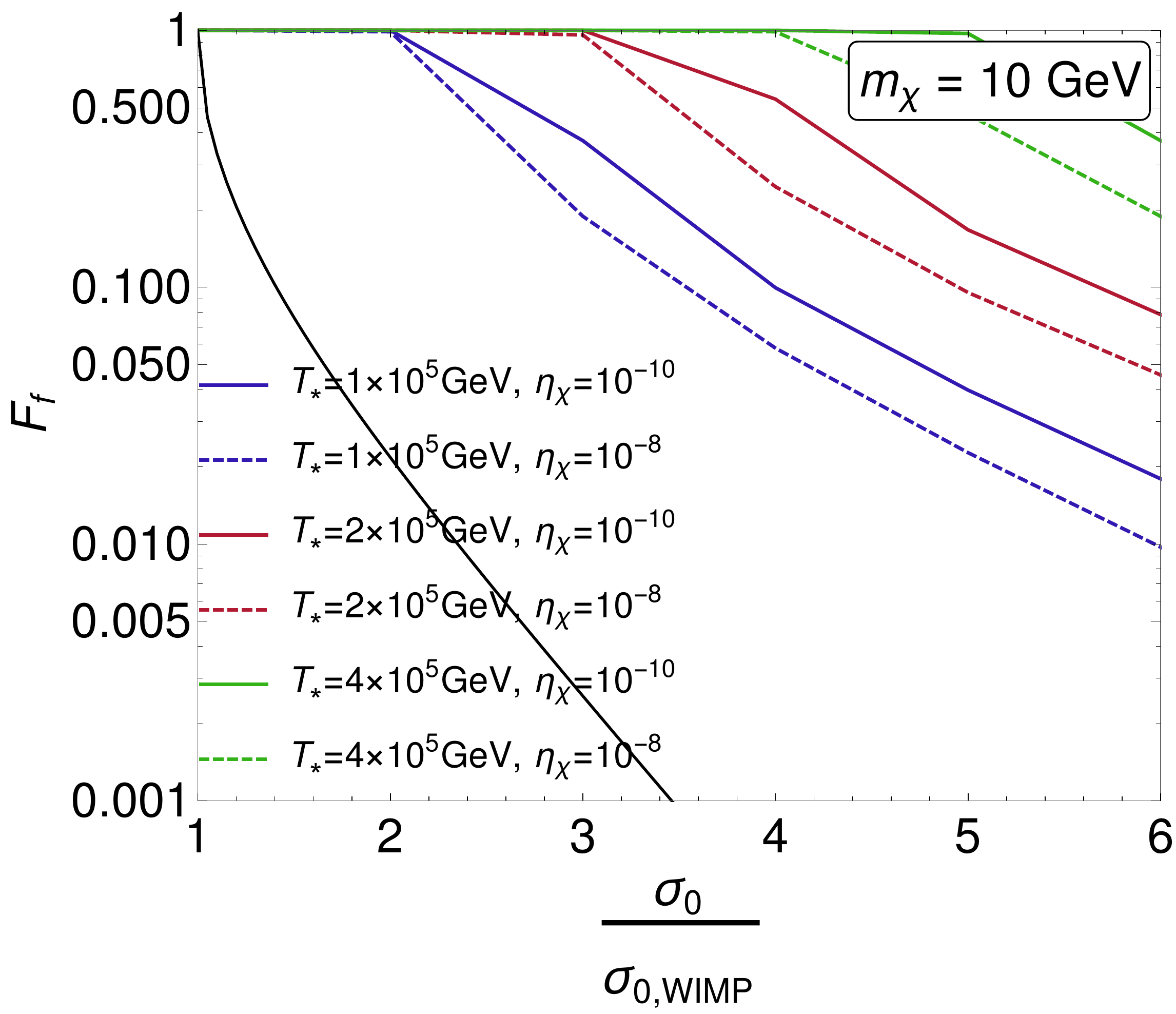}
\includegraphics[width=0.5\textwidth]{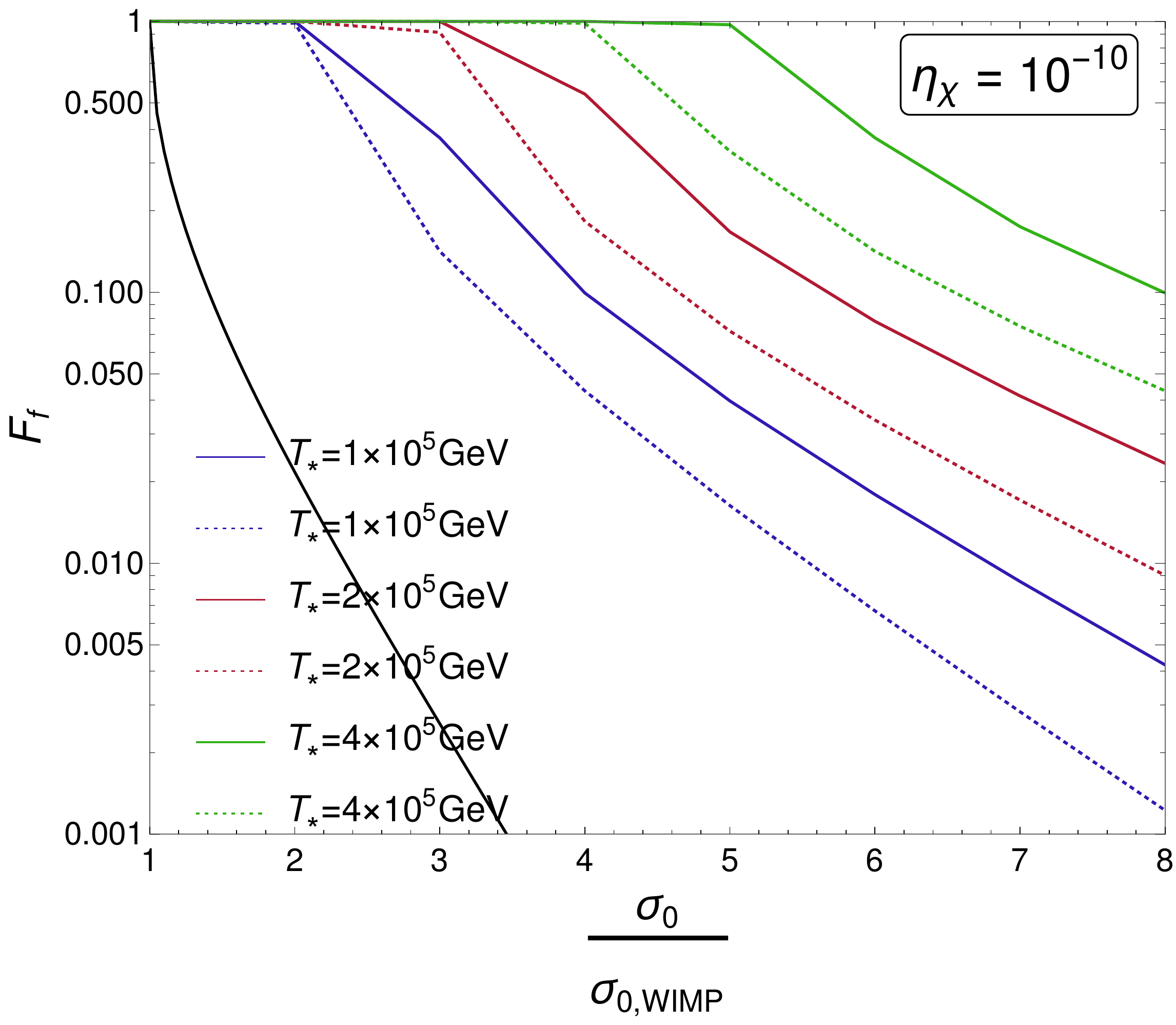}}
\vspace{-3mm} \caption{The present fractional asymmetry $F_{f}$ as function of the cross-section $\sigma_{0}$ relative to the traditional WIMP cross section $\sigma_{0,{\rm WIMP}}$. One can observe variations of the late time fractional asymmetry $F_{f}$ with $\sigma_{0}/\sigma_{0,{\rm WIMP}}$, for a fixed mass (left panel) and  for a fixed asymmetry $\eta_{\chi}$ (right panel) for two different dark matter masses  $m_{\chi} = 1.0\times 10^{4}~{\rm GeV}$ (solid line) and $m_{\chi} = 1.5\times 10^{4}~{\rm GeV}$ (dashed line). The case of decoupling in the radiation dominated era is shown as the black curve.
}
\label{fig:FinfSigmaplot}
\end{figure}

Conversely, we can look to ascertain the annihilation cross section appropriate to reproduce the correct relic density.
 Again, we emulate the treatment of \cite{Graesser:2011wi} and give our results in terms of 
 the thermal WIMP cross section $\sigma_{0,\rm WIMP}$ which is appropriate for giving the correct relic density for symmetric s-wave ($n=0$) freeze-out  in a radiation dominated era. We write this  in terms of the  ratio  $\Omega_{\chi}/\Omega_{B}\simeq 5.5$ and $\Phi_{0,\rm WIMP} = x_{f}^{-1}g_*^{1/2}$ (following from eq.~\eqref{Phi})
  \beq\label{eq:sigmaWIMP}
\sigma_{0,{\rm WIMP}}& =\sqrt{\frac{180}{\pi}}\frac{k^{1/2}\Phi^{-1}_{ 0,{\rm WIMP}}}{m_{p}\eta_{B}M_{\rm Pl}}\frac{\Omega^{\rm Relic}_{B}}{\Omega_{\rm FO}}.
\eeq
Since we can neglect the dependence of $x_{f}$ on $\sigma_{0}$, because of the logarithmic nature of $x_{f}$, we take  $x_{f}\simeq 25$ which implies that the symmetric radiation dominated freeze-out cross section is of order
$  \sigma_{0,{\rm WIMP}} \simeq 2.7\times 10^{-9}  {\rm GeV}^{-2}$. In Figure~\ref{fig:FinfSigmaplot} we show the final fractional asymmetry $F_{f}$ as a  function of cross-section $\sigma_{0}$ relative to the traditional WIMP cross section $\sigma_{0,{\rm WIMP}}$.

When the initial dark matter asymmetry $\eta_{\chi}$ is  relatively small such that $F_{f}\lesssim 1$, the dark matter relic density cannot be determined only with the asymmetry as the second term of the RHS of eq.~\eqref{eq:omega3} gives a significant contribution. Using eqns.~\eqref{lambda}, \eqref{eq:Finf}, \eqref{eq:ObsOmega1},  \& \eqref{eq:sigmaWIMP} one has the following expression involving the fractional asymmetry  \cite{Graesser:2011wi}
\beq\label{eq:Finf1}
F_{f} = \exp\left(-4\nu\frac{\sigma_{0}}{\sigma_{0,{\rm WIMP}}}\left(\frac{1-F_{f}}{1+F_{f}}\right)\frac{\Phi_{0}}{\Phi_{0,{\rm WIMP}}}\right).
\eeq
where we recall that $\nu$ first appears in eq.~(\ref{eq:H3}) and during a matter dominated era $\nu = 2/3$.

\clearpage

Therefore the fractional asymmetry $F_{f}$ can be expressed as a function of the s-wave cross-section $\sigma_{0}$ the case of dark matter freeze-out during matter domination as follows
\beq\label{eq:Finf2}
F_{f} =
\exp\left(-\frac{4}{3}(1-r)^{-1/2}\left(\frac{x_{\star}}{x_{f}}\right)^{1/2}\frac{\sigma_{0}}{\sigma_{0,{\rm WIMP}}}\left(\frac{1-F_{f}}{1+F_{f}}\right)\right)~.
\eeq
Notably, the strong dependence of $F_{f}$ on the s-wave cross-section $\sigma_{0}$, can be seen in  Figure~\ref{fig:FinfSigmaplot}.  Observe that as the final fractional asymmetry $F_f$ is reduced then the cross section $\sigma_0$ must be increased to obtain the correct relic abundance, this is because a reduction in $F_f$ implies the dark matter is more asymmetric and thus in order to remove the symmetric component of the dark matter pair annihilations must be more efficiently. Moreover, since it is assumed that the asymmetry fixes the relic density this is equivalent to avoiding dark matter overproduction.


 \section{Asymmetric Dark Matter from SO(10)}
\label{Sec2}

As a concrete example we highlight here the potential candidates for asymmetric dark matter which arise in the context of non-supersymmetric SO(10) Grand Unified Theories. 
An elegant connection between ADM models and SO(10) GUTs was recently highlighted in \cite{Nagata:2016knk},\footnote{For alternative discussions of ADM candidates within non-SUSY SO(10) models see \cite{Kadastik:2009cu,Kadastik:2009dj,Mambrini:2013iaa,Mambrini:2015vna,Brennan:2015psa,Nagata:2015dma,Arbelaez:2015ila,Boucenna:2015sdg,Mambrini:2016dca,Parida:2016hln}.} in which the lightest member of the scalar multiplets \textbf{16} and \textbf{144}, being a complex singlet with hypercharge zero, is identified as the dark matter state $\chi$. A stabilizing $Z_{2}$ symmetry emerges for $\chi$ from the intermediate scale breaking of a subgroup by a \textbf{126} dimensional representation \cite{Kibble:1982ae,Krauss:1988zc,Ibanez:1991hv,Ibanez:1991pr,Martin:1992mq}. 
As discussed in \cite{Nagata:2016knk}, alternatives potential dark matter candidates from the SO(10) representations (other than the singlet state $\chi$), either disrupt gauge coupling unification or, in the case of non-zero hypercharge states, encounter strong direct detection limits.\footnote{In what follows we  explore how the experimental limits on the singlet state can be relaxed due to an entropy injection from RH neutrino decays and, plausibly, these candidates might be similarly salvaged.}

\subsection{Higgs Portal to Singlet Dark Matter from SO(10)}

While in \cite{Nagata:2016knk}, it was concluded that the singlet state was only a viable ADM state for freeze-out via resonant annihilation to Standard Model states, here we show that if the lightest right-handed (RH) neutrino $N_1$ decays after dark matter freeze-out, then this significantly widens the viable parameter space.  This presents an excellent example of matter dominated freeze-out, with the RH neutrino dominating the energy density for a period of the early~universe.

The dark matter $\chi$ is assumed to have couplings such that it is initially in thermal equilibrium with the Standard Model and then subsequently freezes out non-relativistically. We assume that $\chi$ has a particle asymmetry, for a discussion of the origins of this particle asymmetry in the context of the SO(10) models, see \cite{Nagata:2016knk}. Since $\chi$ is a complex singlet with a $Z_2$ symmetry, it interacts with the Standard Model through renormalizable interactions involving the Higgs, the so called Higgs portal, with the following Lagrangian  \cite{Silveira:1985rk,McDonald:1993ex,Burgess:2000yq,Patt:2006fw} (and see \cite{Arcadi:2021mag} for a recent review)
\beq\label{eq:HportalScalarDM}
 \mathcal{L} = \mathcal{L}_{\text{SM}}+\frac{1}{2}(\partial_{\mu}\chi)^\dagger\partial^{\mu}\chi -\frac{1}{4!}\lambda'|\chi|^{4}-\frac{1}{2}\mu'{}^{2}|\chi|^{2}-\kappa |\chi|^{2}|H|^2,
\eeq
  where $\mathcal{L}_{\text{SM}}$ is the Standard Model Lagrangian.  After electroweak symmetry breaking, the Lagrangian eq.~\eqref{eq:HportalScalarDM} can be expanded around the Higgs VEV to obtain 
  \beq\label{eq:HportalScalarDMexpand}
 \mathcal{L} = \mathcal{L}_{{\rm SM}}+\frac{1}{2}(\partial_{\mu}\chi)^\dagger\partial^{\mu}\chi -\frac{1}{4!}\lambda'|\chi|^{4}-\frac{1}{2}m_\chi^{2}|\chi|^{2}-\frac{1}{2}\kappa |\chi|^{2}h^{2}-\kappa vh|\chi|^2~ .
\eeq
The state $\chi$ acquires mass contributions from both the $\mu'$ term and the cross-coupling term, such that the dark matter mass is $m_{\chi} = \sqrt{\mu'{}^{2}+\kappa v^{2}}$, where $v$ is the Higgs VEV. For $m_{\chi}>10$ GeV and $\kappa<0.1$ (which is where we shall focus). The dark matter mass is set via $m_{\chi}\simeq \mu'$ to a good approximation. This Lagrangian fixes the interactions between the dark matter and Standard Model and, in particular, one can calculate the annihilation cross-section of $\chi$ to Standard Model states. The dark matter will primarily annihilate to Standard Model fermions, gauge bosons, or Higgs bosons depending on which are kinematically accessible (i.e.~depending on $m_\chi$), and we state the main annihilation cross-sections in Appendix \ref{AA}.

As previously, it will be helpful to consider the fractional asymmetry $F = Y_{\chi^{\dagger}}/Y_{\chi}$ and we will express  the dark matter relic abundance  in terms of $\eta_{\chi}$ and $F_f$ according to eq.~(\ref{eq:omega3}) which we recall below
\beq
\Omega^{\rm Relic}_{\rm DM}
 =\zeta\left( \frac{s_{0}m_{\chi}}{\rho_{c}}\right)\left(\eta_{\chi}+\frac{2\eta_{\chi}F_{f}}{1-F_{f}}\right),
\eeq
where $\zeta$ is a dilution factor due to the entropy injection from the decays of $N_{1}$ states.

\subsection{Right-Handed Neutrino Dominated Era}

  The RH neutrino states couple with the Higgs through the Yukawa coupling $y$, with the following Lagrangian
\beq
\mathcal{L}\supset i\bar{N}_{i}\slashed{\partial}N_{i} -\left[y_{ik}\bar{\ell}_{i}HN_{k} +M_{ik}N_{i}N_{k} \right],
\eeq
to generate masses for the left-handed neutrinos through the seesaw mechanism, where $\ell_{i}= e,\mu,\tau$ represents the lepton doublets, $H$ is the Higgs doublet, and $N_{i}$ stands for the RH neutrino states. The decay rate of the RH neutrinos is thus
\beq\label{gaga}
\Gamma_{N_{i}} = \frac{y_{i}^{2}}{8\pi}M_{N_{i}}.
\eeq
Since empirically, the lightest active neutrino can be arbitrarily light (or even massless) \cite{Tanabashi:2018oca}, the neutrino Yukawa matrix can exhibit substantially smaller couplings for one generation of the RH neutrinos $N_{1}$, compared to $N_{2}$ and $N_{3}$. Since the mass of the lightest RH neutrino, $M_{N_{1}}$, can be considerably lower, this can result in a long-lived heavy particle species. Then the population of $N_1$ states can potentially evolve to dominate the energy density of the universe, leading to a matter dominated phase of the early universe  until $N_1$ decays  \cite{Randall:2015xza}. In Appendix \ref{A2} we confirm that for appropriate parameter choices that $N_1$ can indeed lead to an early period of matter domination, for instance with  $\kappa\sim 10^{-12}$ and $M_{N_{1}}\sim 10^{10}$ GeV.

Assuming that the $N_1$ states decay simultaneously at  $H\simeq \Gamma_{N_{1}}$ into Standard Model radiation the resulting reheat temperature $T_{\rm RH}$ for the Standard Model thermal bath will be
\beq
T_{\rm RH} 
= \left(\frac{45 M_{\rm Pl}^{2}}{256\pi^{5} g_*}\right)^{1/4}y_{1}M_{N_{1}}^{1/2}
\simeq 
20~{\rm GeV}
\left(\frac{45 M_{\rm Pl}^{2}}{256\pi^{5} g_*}\right)^{1/4}
\left(\frac{y_{1}}{10^{-12}}\right)
\left(\frac{M_{N_{1}}}{10^{10}~{\rm GeV}}\right)^{1/2}.\label{222}
\eeq
Then the dilution $\zeta$ of the relic abundance due to an entropy injection from $N_1$ decays to Standard Model states can be found from the reheat temperature $T_{\rm RH}$ of the thermal bath following $N_1$ decays as given in eq.~(\ref{000}), which we restate below for convenience
\beq
\zeta  \simeq \frac{45}{4\pi^{3}}   \frac{T_{\rm RH} }{(1-r)T_{\star}g_*}~,
\eeq
where $T_\star$ is the value of the temperature of the Standard Model bath at the point that the $N_1$ become non-relativistic (characteristically $M_{N_1}$) and $r$ is the fraction of the total energy density of the universe in $N_1$ at $T=T_\star$. 

\begin{figure}[t!]
\centerline{\includegraphics[scale=0.35]{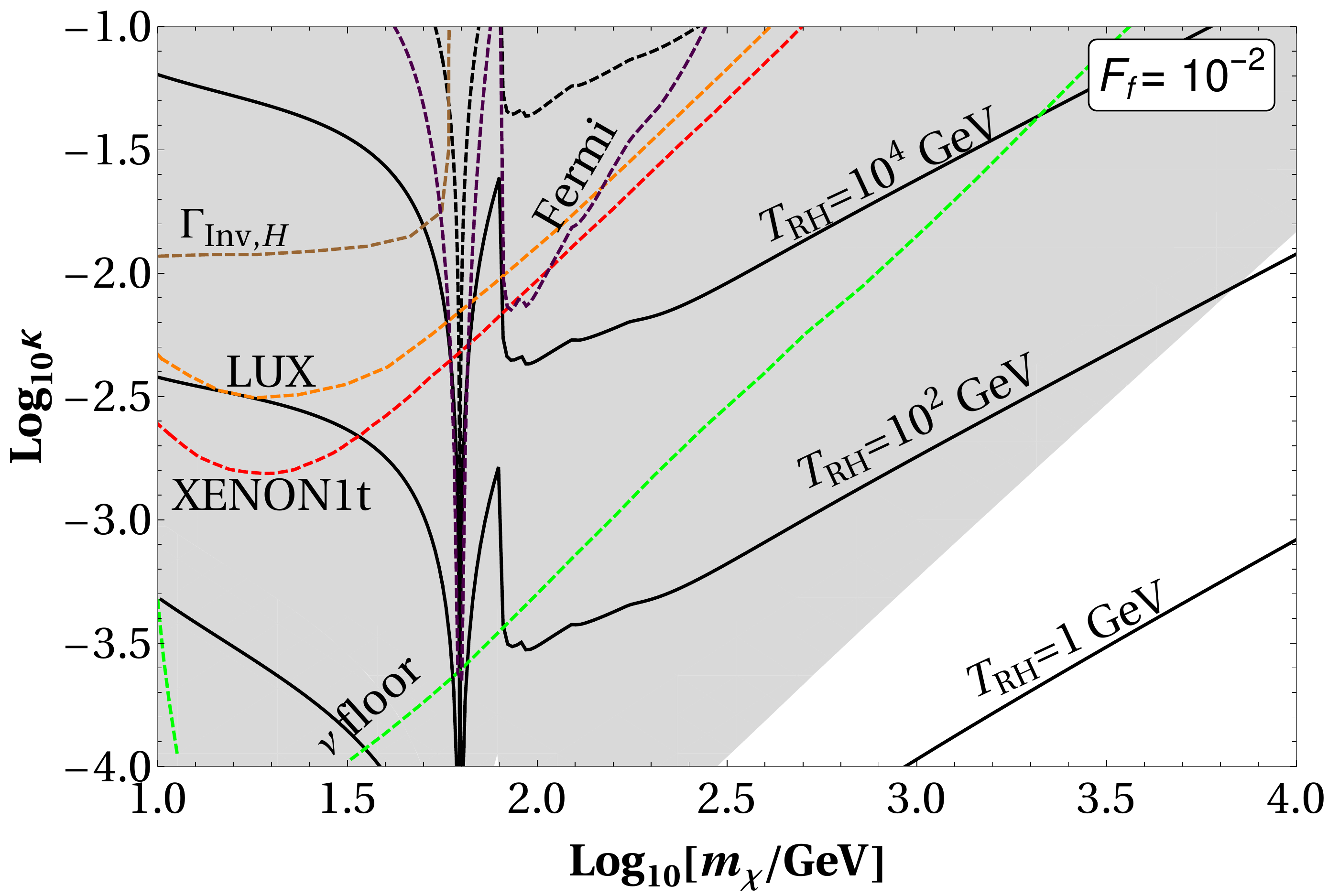}}
\vspace*{-5mm}
\caption{An entropy injection dilutes the dark matter after freeze-out introducing a new parameter for calculating the dark matter relic density, the energy from the entropy injection sets the reheat temperature of the Standard Model radiation bath to $T_{\rm RH}$. We show contours of $T_{\rm RH}$ for which $F_f=Y_{\overline{\rm DM}}/Y_{\rm DM}=10^{-2}$ evaluated at freeze-out, thus the dark matter is asymmetric and reproduces the observed  relic density for couplings on or above a given contour. The model under consideration is a complex scalar dark matter annihilating through the Higgs portal, as motivated by the SO(10) model discussed here. Experimental constraints are shown from XENON1T \cite{Aprile:2018dbl} (dashed red), LUX \cite{Akerib:2016vxi} (dashed orange), Fermi-LAT \cite{Ackermann:2015zua} (dashed purple) and the invisible Higgs width \cite{Escudero:2016gzx, Khachatryan:2016whc} (dashed brown). The grey shaded region indicates where the theoretical analysis breaks down, corresponding to the assumption of a matter dominated universe. The neutrino floor is shown as the dashed green curve. We assume that the onset of matter domination occurs at $T_{\rm MD} = M_N/100\sim 10^{8}~{\rm GeV}$, as might arise for $T_\star\sim M_N$  with the fraction of energy in $N_1$ at $T=T_\star$ being $r=0.01$, cf.~eq.~(\ref{MDD}). The white space in each panel is viable parameter space in which the dark matter relic density is correct while evading experimental limits for an appropriate choice of $T_{\rm RH}$ (it is, however, below the neutrino floor). We note that reducing $F_f$ by an order of magnitude or two leads to quite similar plots.
\label{fig:HportalKappaDMmass}} 
\end{figure}

\subsection{Widening the Space of Viable SO(10) Scalar Asymmetric Dark Matter}

Taking the above together, in Figure \ref{fig:HportalKappaDMmass}  we present the parameter space for SO(10) singlet dark matter $\chi$ which freezes-out during an early matter dominated era due to decoupled $N_1$ states with mass $M_{N_1}\sim10^{10}$ GeV dominating the energy density of the universe. The contours show the couplings required to reduce the symmetric component of $\chi\chi^{\dagger}$-pairs such that the fractional asymmetry at freeze-out is $F_f=10^{-2}$ for different values of the reheating temperature. The initial dark matter asymmetry is adjusted to match the observed relic abundance as $m_\chi$ and $T_{\rm RH}$ are varied, given that $\Omega\propto \zeta m_{\chi}\eta_{\chi}\big|_{\rm initial}$.  The grey region indicates regions in which our assumptions breakdown, namely in which dark matter freeze-out occurs prior to $N_{1}$ states dominating the energy density or near the point of $N_1$ decays around $H\sim\Gamma$.
To match observations the decays of $N_{1}$ should produce a  reheating temperature higher than the threshold of BBN at around 10 MeV \cite{Sarkar:1995dd,Khlopov}.  For more discussion on these conditions, see \cite{Chanda:2019xyl} which presents similar plots and has additional details.

We overlay Figure \ref{fig:HportalKappaDMmass}  with experimental constraints  from XENON1T \cite{Aprile:2018dbl}, LUX \cite{Akerib:2016vxi} Fermi-LAT \cite{Ackermann:2015zua} (dashed purple) and LHC determinations of the invisible Higgs width \cite{Escudero:2016gzx, Khachatryan:2016whc} (dashed brown), as well as the predicted neutrino floor. Observe that for the scenario presented, which assumes a fractional asymmetry of $10^{-2}$, there is viable parameter space in which $\chi$ can be ADM and simultaneously satisfy the relic density requirement and experimental bounds over a broad range of masses for reheat temperatures below 100 GeV. Note that the many orders of magnitude difference between the orthodox WIMP scenario (dashed) and this scenario with $T_{\rm RH} \sim 1$ GeV is mainly due to the fact that the relic abundance is diluted by a factor of $\zeta$, as given in eq.~(\ref{000}). Accordingly, the required coupling can be smaller by a factor $\sqrt{\zeta} \sim 10^{-6} \sqrt{\frac{T_{\rm RH}}{1~{\rm GeV} }}\sqrt{\frac{10^{10} {\rm GeV} }{M_N}}$.
Note that while it is unfortunate that the parameter  space lies  below the neutrino floor, these are perfectly acceptable dark matter scenarios which could be readily realised in nature if the UV completion of the Standard Model is a non-supersymmetric SO(10) GUT.


\vspace{-2mm}
\section[Concluding Remarks]{Concluding Remarks: Superheavy Asymmetric Dark Matter}
\label{Sec5}
\vspace{-2mm}

In closing, it is interesting to note that this class of models provides a potential route to realising Superheavy Asymmetric Dark Matter (where we take `superheavy' to mean dark matter with mass 1 PeV $\lesssim m_{\chi}\lesssim M_{\rm Pl}$). 
The conventional motivation for ADM is that it offers an explanation for the observed $\mathcal{O}(1)$ coincidence of the present day baryon and dark matter abundances: $\Omega_{\chi}\approx5\Omega_{B}$. Specifically, if one assumes that $\eta_{\chi}\sim\eta_B$ by linking the baryon and dark matter asymmetries, and the dark matter mass is similar to the proton mass $m_{\chi}\sim m_p\approx 1$ GeV, then 
\begin{equation}
\frac{\Omega_{\chi}}{\Omega_{B}}\simeq \frac{\eta_{\chi}m_{\chi}}{\eta_B m_{\rm p}}\sim\mathcal{O}(1)~.
\label{eq1}
\end{equation}
However, this relationship also holds  away from the assumptions $\eta_{\chi}/\eta_B\sim m_{\chi}/ m_p\sim1$. Indeed, the conspiracy of scales between the proton and dark matter typically continues to beg for an explanation (although, see Cohen {\em et al.} for \cite{Cohen:2010kn} one possible realization).
Notably, from inspection of Figure \ref{fig:omega1} this framework of matter domination decoupling of ADM provides motivated examples of this less studied ADM scenario with $m_{\chi}\gg m_p$.

Notably, Superheavy ADM models are also partially motivated by a second, distinct, compelling observation, namely that these scenarios can have significant impacts on astrophysical bodies.  Notably, significant quantities of ADM can accumulate in objects such as main sequence stars, white dwarfs, and neutron stars  \cite{Goldman:1989nd,Gould:1989ez,Sandin:2008db,Kouvaris:2010jy,Frandsen:2010yj,McDermott:2011jp}. 
The reason ADM can have a larger impact than conventional `symmetric' dark matter is due to the fact that the present-day abundance of anti-dark matter $Y_{\overline{\chi}}$ is typically negligible in ADM and therefore dark matter pair annihilations are rare even in dense environments.  
In particular, it has been suggested that ADM in the mass range  0.1-100 PeV could be responsible for the collapse of pulsars near the Galactic Center \cite{Bramante:2014zca} and ignition of type-Ia supernovae \cite{Bramante:2015cua,Graham:2015apa,Acevedo:2019gre}, both of which are open problems in astrophysics.

The central obstruction to realizing Superheavy ADM is that for the relic density to be asymmetric, the symmetric component of pairs of dark matter and anti-dark matter states must be removed via annihilations. For heavier dark matter, a smaller asymmetry is required since fewer particles are needed at late time to make up the observed relic abundance. For instance, for PeV mass ADM one requires $\eta_{\chi}^{\rm today}\sim 6 \times 10^{-16}$ and  in accordance with eq.~(\ref{eq1}) this scales as follows
\beq
\frac{\Omega_{\chi}^{\rm Relic}}{\Omega_{B}^{\rm Relic}}\simeq  5\times\left(\frac{m_{\chi}}{1~{\rm PeV}}\right)\left(\frac{\eta_{\chi}^{\rm now}}{6 \times 10^{-16}}\right)~.
\label{omega5}
\eeq
Since the asymmetry is smaller, a larger annihilation rate is needed for $Y_{\overline{\chi}}\ll  Y_{\chi}$ to ensure that the final relic density is asymmetric. It follows that in calculating the annihilation cross section needed to deplete the symmetric component of dark matter for increasing mass, leads to a unitarity limit analogous to the classic Griest and Kamionkowski unitarity bound \cite{Griest:1989wd}. 

Assuming perturbative annihilations, the ADM mass bound is roughly $m_{\chi}\lesssim100$~TeV \cite{Baldes:2017gzw}.
As seen in Section \ref{Sec4} this ADM unitarity bound can be evaded via the introduction of an entropy injection event in the early universe. 
Specifically, perturbative unitarity \cite{Griest:1989wd} constraints the s-wave annihilation cross-section to satisfy 
$  \sigma_{0}\lesssim \frac{4\pi}{m_{\chi}^{2}}$.
For a specific critical temperature $T_{\star} = 10^{5}$ GeV, the upper bound on the dark matter mass from the unitarity limit is deduced from eq.~\eqref{eq:omega5} to be $m_{\chi}\lesssim 10^{11} ~{\rm GeV}$.

The first model of  Superheavy Asymmetric Dark Matter was introduced  in  \cite{Bramante:2017obj} (an alternative class of models was outlined in \cite{Ghosh:2020lma}). 
While the work of  \cite{Bramante:2017obj} provides a proof of principle that Superheavy ADM can give the correct relic density, it introduces a strong assumption regarding the ordering of events and thus does not fully explore the potential parameter space. Specifically, it was assumed that dark matter freeze-out occurred during an era of radiation domination. However, the entropy production, which is assumed to follow dark matter freeze-out, must be sourced by some other significant energy density in the early universe, and this can potentially significantly alter the cosmological history. For instance, if the entropy injection is due to the decays of some heavy decoupled particle species, then over a large swathe of parameter space, the dark matter will freeze-out in a matter dominated universe.  Since this case was neglected in this initial study \cite{Bramante:2017obj}, this work completes this interesting picture by exploring the parameter space of Superheavy ADM which decouples during matter domination.
\\

\noindent
{\bf Acknowledgements.} We are grateful to Jakub Scholtz for helpful comments.

\appendix

\section{Appendices}

\vspace{2mm}
\subsection{Entropy Production in the Thermal Bath During Decoupling}
\vspace{2mm}\label{APD} 

    While the assumption that $\phi$ states decay simultaneously at $H\simeq\Gamma_\phi$ provides a useful simplification, the natural expectation is a exponential decay law for the $\phi$ particles \cite{Scherrer:1984fd} compared to the instantaneous decay of $\phi$ at some $t\sim \Gamma^{-1}$. However, the approximation of sudden decay of the $\phi$ particles remains valid until the entropy violation caused by the radiation due to the decay of the $\phi$ particles can no longer be neglected, such that the old radiation becomes comparable to the new radiation produced by $\phi$ decay \cite{Scherrer:1984fd}. 
    
  One can define a temperature  $T_{\rm EV}$ \cite{Chanda:2019xyl}, so that for $T\gtrsim T_{\rm EV}$ the universe remains matter dominated $H\propto T^{3/2}$ and $T\propto a^{-1}$ as the $\phi$ decays are unimportant, whereas for $T\lesssim T_{\rm EV}$ the $\phi$ decays become significant such that $H\propto T^{4}$ and $T\propto a^{-3/8}$ as in \cite{Scherrer:1984fd}. This temperature threshold at which entropy violation in the bath is non-negligible is given by 
\beq\label{eq:TEV}
T_{\rm EV} \simeq T_{\star}\left(1+\left(\frac{r}{1-r}\frac{\mathrm{v}}{\Gamma t_{\star}}\right)\right)^{(1-\mathrm{v})/\mathrm{v}},
\eeq    
    where $\mathrm{v}\equiv 2/3(1+\omega)^{-1}+1$, with $\omega = 0$ and 1/3 implies matter domination and radiation domination respectively, and $t_{\star}$ is defined as the time when $\phi$ starts behaving matter-like ($t_{\star}\sim H_{\star}$). Thus, for dark matter freeze-out to occur during matter domination, it is required that $T_{f}\gtrsim T_{\rm EV}$. If, however, $\phi$ decays become non-negligible before the dark matter decouples from the thermal bath, the dark matter freeze-out calculations should be performed similar to the classic paper of Giudice, Kolb, \& Riotto \cite{Giudice:2000ex} and we do not consider this case here.

\subsection{Derivation of the Asymmetric Yield}
\label{AnalyticYX}

  Here  an analytic solution of Boltzmann equation eq.~\eqref{eq:be3} is obtained with the similar procedure as described in \cite{Iminniyaz:2011yp} to arrive at eq.~(\ref{eq:ADMabundance0}) in the main body.  Defining $ \Delta^{-}= Y^{-}-Y^{-}_{\rm{\rm eq}}$ the anti-DM abundance is given by eq.~\eqref{eq:be3} 
\beq\label{eq:analytic2}
 \frac{dY^{-}}{dx}& = -\sum_{n}\lambda_{n} x^{-n-2}g_{\rm eff}^{1/2}(\Delta^{-}(\Delta^{-}+2Y^{-}_{\rm{\rm eq}})+\eta_{\chi} \Delta^{-})\\
 \frac{d\Delta^{-}}{dx}& =-\frac{dY^{-}_{\rm{\rm eq}}}{dx} -\sum_{n}\lambda_{n} x^{-n-2}g_{\rm eff}^{1/2}(\Delta^{-}(\Delta^{-}+2Y^{-}_{\rm{\rm eq}})+\eta_{\chi} \Delta^{-})~,
\eeq
where we have used  $Y_{\rm{\rm eq}}^{+}-Y_{\rm{\rm eq}}^{-}= \eta_{\chi}$.
  Much before the freeze out, when the temperature was  high ($1<x\ll x_{f}$), $Y^{\pm}$ tracks $Y^{\pm}_{\rm{\rm eq}}$ very closely.  During that period of time, $\Delta^{-}$ and $\frac{d\Delta^{-}}{dx}$  become very small. Neglecting $\frac{d\Delta^{-}}{dx}$ and higher order terms of $\Delta^{-}$ in eq.~\eqref{eq:analytic2} we obtain
\beq\label{analytic4}
\frac{dY^{-}_{\rm eq}}{dx} \simeq -\sum_{n}\lambda_{n} x^{-n-2}g_{\rm eff}^{1/2}(2\Delta^{-}Y^{-}_{\rm{\rm eq}}+\eta_{\chi} \Delta^{-}).
\eeq
 At equilibrium $\frac{dY^{-}_{\rm eq}}{dx}$ vanishes, then from eq.~\eqref{eq:be3} and eq.~\eqref{F2}, one has 
\beq\label{eq:analytic5}
&Y_{\rm eq}^{-}(Y_{\rm eq}^{-}+\eta_{\chi})-Y_{\rm{\rm eq}}^{2}=0  \qquad {\rm and} \qquad
&   Y_{\rm eq}^{-}=-\frac{\eta_{\chi}}{2}+\sqrt{\frac{\eta_{\chi}^{2}}{4}+Y_{\rm{\rm eq}}^{2}} .
\eeq
   Inserting eq.~\eqref{eq:analytic5} into eq.~\eqref{analytic4} we get
\beq
\frac{1}{2}\left(\frac{\eta^{2}}{4}+Y_{\rm{\rm eq}}^{2}\right)^{-1/2}\frac{dY^{2}_{\rm{\rm eq}}}{dx} = -\sum_{n}\lambda_{n} x^{-n-2}g_{\rm eff}^{1/2}\Delta^{-}\left(2\sqrt{\frac{\eta_{\chi}^{2}}{4}+Y_{\rm{\rm eq}}^{2}}\right).
\eeq
Then using eq.~\eqref{eq:F3}  and restricting to 
the epoch of  interest $1<x\ll x_{f}$,  for which case the  $x^{2}$ term can be neglected  compared to the  $x^{3}$ term, we find that
\beq
 b^{2}(-2x^{3})e^{-2x}\simeq -4\sum_{n}\lambda_{n} x^{-n-2}g_{\rm eff}^{1/2}\Delta^{-}\left(\frac{\eta_{\chi}^{2}}{4}+Y_{\rm eq}^{2}\right),
\eeq
which leads us to the following solution 
\beq
\Delta^{-}\simeq \frac{Y_{\rm{\rm eq}}^{2}}{2g_{\rm eff}^{1/2}\sum_{n}\lambda_{n} x^{-n-2}\left(\frac{\eta_{\chi}^{2}}{4}+Y_{\rm eq}^{2}\right)}.
\eeq

  At later times much after the freeze out, the temperature will be sufficiently low ($x\gg x_{f}$) such that  $Y^{-}$ tracks $Y_{\rm{\rm eq}}^{-}$ very poorly. As a result, the abundance $Y^{-}$ becomes comparable to $\Delta^{-}$ ($\Delta^{-}\simeq Y^{-}\ll Y_{\rm{\rm eq}}^{-}$). This allows us to neglect terms in eq.~\eqref{eq:analytic2}, which depends on $\frac{dY_{\rm{\rm eq}}^{-}}{dx}$ and $Y_{\rm{\rm eq}}^{-}$, leading to the expression
\beq\label{eq:analytic3}
\frac{d\Delta^{-}}{dx}\simeq -\sum_{n}\lambda_{n} g_{\rm eff}^{1/2}x^{-n-2}(\Delta^{-2}+\eta_{\chi}\Delta^{-}).
\eeq
From the above we can obtain $\Delta^{-}$ by integrating in the interval $[x_{f},\infty]$ 
 \beq
 \int_{x_{f}}^{\infty}\frac{d\Delta^{-}}{(\Delta^{-2}+\eta_{\chi}\Delta^{-})}=-\int_{x_{f}}^{\infty}dx\sum_{n}\lambda_{n} g_{\rm eff}^{1/2}x^{-n-2},
 \eeq
and partially evaluating yields
\beq
\frac{1}{\eta_{\chi}}\ln \frac{\Delta^{-}}{\Delta^{-}+\eta_{\chi}}\Big|_{x_{f}}^{\infty} = -\sum_{n}\int_{\bar{x}_{f}}^{\infty}dx~\lambda_{n} g_{\rm eff}^{1/2}x^{-n-2}.
\eeq
For $x\gg x_{f}$, one can assume $\Delta^{-}(x_{f})\gg \Delta^{-}_\infty $, which leads us to the following expression
\beq
   \frac{1}{\eta_{\chi}}\ln\left(\frac{\Delta^{-}_\infty +\eta_{\chi}}{\Delta^{-}_\infty }\right)\simeq \sum_{n}\int_{x_{f}}^{\infty}dx~\lambda_{n} g_{\rm eff}^{1/2}x^{-n-2}.
 \eeq
It follows that the anti-dark matter abundance set by the decoupling is
\beq
Y_{\infty}^{-}\simeq \Delta_{\infty}^{-} = \frac{\eta_{\chi}}{\exp\left(\eta_{\chi}\sum\limits_{n}\int\limits_{x_{f}}^{\infty}dx~\lambda_{n} g_{\rm eff}^{1/2}x^{-n-2}\right)-1},
\eeq


Furthermore, since the dark matter mass $m_{\chi}$ is independent of the bath temperature $T$, the parameter $\lambda_{n}$ can be pulled out of the integration and the anti-dark matter abundance can be further simplified to obtain
\beq
Y_{\infty}^{-} \simeq \frac{\eta_{\chi}}{\exp\left(\eta_{\chi}\sum\limits_{n}\lambda_{n}\Phi_{n}(\infty ,m_{\chi})\right)-1},
\eeq
with the definition
$\Phi_{n} (x,m_{\chi}) = \int_{x_{f}}^{x} dx^{\prime} x^{\prime -n-2}g_{\rm eff} ^{1/2}$.

\subsection{Dark Matter Production in the Transition from Matter Domination to Radiation Domination} \label{DMprodLoop}
 
 While one could be concerned that even in the case that there is negligible production of dark matter from decays of $\phi$, even a small amount of anti-dark matter production might alter $F$ the fractional asymmetry.\footnote{We remind the reader that $\phi$ is some general field and should not be taken to be the inflaton.} We next argue that this is not the case in the parameter regions of interest
 Emulating the treatment of \cite{Kaneta:2019zgw}, the ratio of the contribution to the dark matter relic density from the $\phi$ decay, $\Omega_{\chi, {\rm decay}}$, to that produced from freeze-out, $\Omega_{\rm relic}$, is derived as
 \beq
 \frac{\Omega_{\chi, {\rm decay}}}{\Omega_{\rm relic}} \sim 1.6\times 10^{-6}
 \left(\frac{\mathcal{B}_{\chi}}{4\times 10^{-9}}\right)\left(\frac{10^{10}~{\rm GeV}}{m_{\phi}}\right)\left(\frac{T_{\rm RH}}{1 ~{\rm GeV}}\right)\left(\frac{m_{\chi}}{600{\rm GeV}}\right),
 \eeq
 which can be calculated using the branching ratio derived in \cite{Chanda:2019xyl} 
 \beq
 \mathcal{B}_{\chi}\sim 4\times 10^{-9}\left(\frac{\kappa}{0.01}\right)^{2}.
 \eeq
It follows that, for the selected parameter space above,  $\Omega_{\chi, {\rm decay}}$ is obtained as $\sim 10^{-6}$, which  is much smaller compared to the density of dark matter and anti-dark matter produced through the freeze-out which are normalized, such to agree with the observed relic density.

\subsection{Higgs Portal Annihilation Cross Sections}
\label{AA}

In this appendix we give the relevant annihilation cross sections for Higgs Portal dark matter. As can be inferred from eq.~\eqref{eq:HportalScalarDMexpand},  scalar dark matter can annihilate to Standard Model particles via three main routes, (i).~dark matter annihilations to Standard Model fermions $\chi\chi\rightarrow f\bar{f}$; (ii).~dark matter to Standard Model vector bosons $\chi\chi\rightarrow V\bar{V}$; (iii).~dark matter to Higgs bosons $\chi\chi\rightarrow hh$. Accordingly, the  partial cross-sections in terms of the Mandelstam variable $s$ are respectively 
 \beq
 \sigma_{\chi\chi\to f\bar{f}} &= \frac{N_{c}\kappa^{2}m_{f}^{2}}{2\pi s}\frac{\sqrt{1-\frac{4m_{f}^{2}}{s}}}{\sqrt{1-\frac{4m_{\chi}^{2}}{s}}}\frac{s-4m_{f}^{2}}{(s-m_{h}^{2})^{2}+m_{h}^{2}\Gamma_{h}^{2}}\\
 \sigma_{\chi\chi\to V\bar{V}} & =\frac{\delta_{V}\kappa^{2}m_{V}^{4}}{\pi s}\frac{\sqrt{1-\frac{4m_{V}^{2}}{s}}}{\sqrt{1-\frac{4m_{\chi}^{2}}{s}}}\frac{\left(2+\frac{\left(s-2m_{V}^{2}\right)^{2}}{4m_{V}^{4}}\right)}{\left[(s-m_{h}^{2})^{2}+(m_{h}\Gamma_{h})^{2}\right]}\\
 \sigma_{\chi\chi\to hh} &= \frac{\kappa^{2}}{8\pi s}\frac{\sqrt{1-\frac{4m_{h}^{2}}{s}}}{\sqrt{1-\frac{4m_{\chi}^{2}}{s}}} ~,
 \eeq
where $N_{c}$ is the number of colors, $V=Z,W^\pm$ with $\delta_{Z} = \frac{1}{2}$ and $\delta_{W^{+},W^{-}}=1$, and the Higgs boson mass and  width are  $m_h$ and $\Gamma_h\approx13~\mathrm{MeV}$ \cite{Tanabashi:2018oca}.

 Taking the non-relativistic limit, we expand the partial thermally averaged cross-sections $\langle\sigma v\rangle_i$ to order $v^{2}$  (in terms of the relative velocity $v\approx v_{i}/2$ with $v_{i}$ the velocity of a dark matter particle during freeze-out) leading to 
 \beq
 \langle\sigma v\rangle=\sigma_0+\sigma_1 v^{2}+\cdots
 \eeq
Thus  we obtain the following partial annihilation cross sections \cite{Chanda:2019xyl,Lerner:2009xg,Guo:2010hq,Cline:2013gha})
 \beq\label{eq:sDMannH}
 \langle\sigma v\rangle_{\chi\chi\to f\bar{f}} &=
 \frac{N_{c}\kappa^{2}r_{f}^{2} }{4\pi m_{\chi}^{2}}\frac{\left(1-4r_{f}^{2}\right)^{3/2}}{\left(1-r_{h}^{2}\right)^{2}+r_{\Gamma}^{2}}\left[1
+\frac{v^2}{2}\frac{7r_{f}^{2}+r_{h}^{2}(1-10r_{f}^{2})+3r_{f}^{2}(r_{h}^{4}+r_{\Gamma}^{2})-1}{\left(1-4r_{f}^{2}\right)\left[\left(1-r_{h}^{2}\right)^{2}+r_{\Gamma}^{2}\right]}\right]\\[5pt]
 \langle\sigma v\rangle_{\chi\chi\to V\bar{V}} &=
 \frac{\delta_{V}\kappa^{2}}{8\pi  m_{\chi}^{2}}\sqrt{1-4r_{V}^{2}}
 \frac{(1-4r_{V}^{2}+12r_{V}^{4})}{\left(1-r_{h}^{2}\right)^{2}+r_{\Gamma}^{2}} \left[ 1
 +\frac{v^2}{4} \frac{K-1}{\left(1-4r_{V}^{2}+12r_{V}^{4}\right)\left[(1-r_{h}^{2})^{2}+r_{\Gamma}^{2}\right]}\right]\\[5pt]
  \langle\sigma v\rangle_{\chi\chi\to hh}  &=\frac{\kappa^{2}}{16\pi m_{\chi}^{2}}\sqrt{1-4r_{h}^{2}}\left[1
 +\frac{v^2}{4}\frac{6r_{h}^{2}-1}{1-4r_{h}^{2}}\right]~,
 \eeq
in terms of the ratios
 $r_{i} = m_{i}/(2m_{\chi})$  and  $r_{\Gamma}=m_{h}\Gamma_{h}/(4m_{\chi}^{2})$,
 and where $F$ is given by
 \beq
 K=14r_{V}^{2}-76r_{V}^{4}+168r_{V}^{6}-r_{h}^{2}(12r_{V}^{2}-96r_{V}^{4}+240r_{V}^{6})+(r_{h}^{4}+r_{\Gamma}^{2})(1-2r_{V}^{2}-20r_{V}^{4}-72r_{V}^{6}).
\notag
 \eeq 
 By replacing, $v^{2} = 6/x_{f}$ the expressions above can be re-written in terms of the mass scaled inverse temperature $x$.

 \subsection{Early Matter Domination from RH Neutrinos}
\label{A2}

In this section we confirm that for suitable parameters the lightest RH neutrino $N_1$ can lead to an early period of matter domination.
We first demonstrate that if the $N_1$ decay rate is controlled by a coupling of order $y_1\sim 10^{-12}$ then the RH neutrinos are not ever coupled to the Standard Model bath. The requirement that the $N_1$ are not in thermal equilibrium with the Standard Model bath is different for $T>M_{N_{1}}$ and $T<M_{N_{1}}$, as in the latter case the equilibrium abundance is Boltzmann suppressed \cite{Elahi:2014fsa}, so it is convenient to treat these cases separately. Subsequently, we identify the minimum contribution of $N_1$ to the energy density of the universe, as controlled by $r$, in order for $N_1$ domination to occur prior to $N_1$ decays.

 Let us suppose that the population of $N_1$ states as a whole are in internally thermal equilibrium with some temperature $T_1$, which may be different to the temperature $T$ of the Standard Model bath\footnote{We shall assume that the Standard Model states were in thermal equilibrium since $T\sim 10^{16}$ GeV,  it is less obvious whether the Standard Model was ever in equilibrium with a temperature in excess of this.}  and that $T_1\leq T$. A simple manner to ensure that the $N_1$  thermalise is to have a state $\varphi$ with modest couplings to the $N_1$, then subsequently the $\varphi$ VEV can be identified with mass $\langle\varphi\rangle=M_1$ of $N_1$. It then follows that $T_1/T\simeq (\rho_{N_1}/\rho_{\rm SM})^{1/4}$ where $\rho_{N_1}$ and $\rho_{\rm SM}$ are the energy densities of $N_1$ and the Standard Model, respectively. These energy densities are set by inflationary reheating according to the branching ratios of the inflaton. We will take $\xi\equiv T_1/T$ to be a free parameter with $\xi<1$.

 During  early cosmological times, when $T_1\gg M_{N_{1}}$, the $N_1$ interaction rate is 
\beq
\Gamma \sim n_{N_{1}}\langle\sigma v\rangle \sim \frac{y_{1}^{2}T_1}{\pi^{2}}~,
\eeq
where the $N_{1}$ number density $n_{N_{1}}\sim T_1^{3}/\pi^{2}$ and the cross-section $\langle\sigma v\rangle\sim y_{1}^{2}/T_{1}^{2}$, and $y_{1}$ is the Yukawa coupling $y_{1}H\bar{e}N_{1}$. Therefore, we can express the ratio between the interaction rate and the Hubble expansion rate as
\beq
\frac{\Gamma}{H} = \sqrt{\frac{45}{4\pi^{7}g_*}}M_{\rm Pl}\frac{y_{1}^{2}\xi}{T}
< \sqrt{\frac{45}{4\pi^{7}g_*}} M_{\rm Pl} \frac{y_{1}^{2}}{M_{N_1}}
\simeq
10^{-18} \times \left(\frac{y_{1}}{10^{-12}}\right)^{2}\left(\frac{10^{10}~{\rm GeV }}{M_{N_1}}\right)~.
\eeq
For a choice of $y_{1}\sim 10^{-12}$ and $g_*\sim 100$ this suggests that the interaction rate $\Gamma$ is always less than the Hubble expansion rate $H$, as the universe cools down from $T_1\sim 10^{16}$ GeV to a temperature comparable to the $N_{1}$ mass scale. 
Thus the population of $N_{1}$ is initially out of equilibrium with the thermal bath and for $T>M_{N_{1}}$ the abundance of the $N_{1}$ states is such that $Y_{N_{1}}\ll Y^{\rm eq}$. However, at temperatures $T<M_{N_{1}}$ the equilibrium abundance $Y^{\rm eq}$ is Boltzmann suppressed and to prevent the $N_{1}$ abundance $Y_{N_{1}}$ from becoming comparable to the equilibrium value $Y^{\rm eq}$, it is required that the decoupling temperature for the $N_{1}$ states  (which we denote $T_{N_{1},f}$) must be much larger compared to its mass.

Therefore we should check whether $T_{N_{1},f}\gg M_{N_{1}}$ by inspection of $\Gamma /H$  at late times. For $T_1\ll M_{N_{1}}$ the interaction rate of the $N_{1}$ states is
\beq
\Gamma \sim n_{N_{1}}\langle\sigma v\rangle \sim \frac{y_{1}^{2}}{(2\pi)^{3/2}}\left(\frac{T_1^{3}}{M_{N_{1}}}\right)^{1/2}\exp\left(-M_{N_{1}}/T_1\right),
\eeq
where the $N_{1}$ number density is now $n_{N_{1}}\sim \left(M_{N_{1}}T_1/2\pi\right)^{3/2}\exp\left(-M_{N_{1}}/T_1\right)$. Thus the ratio between the interaction rate and the Hubble  rate for  $T_1\ll M_{N_{1}}$ is exponentially suppressed 
\beq
\frac{\Gamma}{H} \sim A \exp \left(-\frac{M_{N_{1}}}{\xi T}\right)~,
\eeq
where the prefactor is given by
\beq
A=\sqrt{\frac{45}{8\pi^{6}g_*}}\frac{M_{\rm Pl}y_{1}^{2}}{\sqrt{M_{N_{1}}T}\xi^{3/2}}
<
\sqrt{\frac{45}{8\pi^{6}g_*}}\left(\frac{y_{1}^{2}M_{\rm Pl}}{M_{N_{1}}}\right)
\simeq 10^{-17}
\left(\frac{y_{1}}{10^{-12}}\right)^{2}
\left(\frac{10^{10}~{\rm GeV}}{M_{N_{1}}}\right).
\eeq
Since $\Gamma/H<A\ll1$ for  $y_1\sim 10^{-12}$, the $N_{1}$ states  are again found to remain decoupled.

 While we have shown that the RH neutrinos remain decoupled we should also show that they are sufficiently long lived to dominate the energy density of the universe.
 By eq.~(\ref{MDD}) we have that the RH neutrino becomes the dominant energy density for
 \beq
 T_{\rm MD} \simeq \frac{(1-r)}{r} \frac{ M_{N_1}}{\xi} \sim M_{N_{1}}\left(\frac{1-r}{r}\right)^{3/4},
\eeq
where we take that $N_1$ becomes non-relativistic at $T_1\sim M_{N_1}$, implying that $T_\star \sim  M_{N_1}/\xi$. We also use in the latter relation that the fraction of energy density in $N_1$ and the $N_1$ temperature are related, such that $\xi\sim \left(\frac{1-r}{r}\right)^{1/4}$.   We next should compare $ T_{\rm MD}$ to $T_{\Gamma}$. The decay rate of $N_1$ is given by eq.~(\ref{gaga}) and thus (in the spontaneous decay approximation) the $N_1$ decay at a temperature 
\beq
T_\Gamma\sim \left(\frac{45}{256\pi^{5}g_{\star}}\right)^{1/4}y_{1}\sqrt{M_{N_{1}}M_{\rm Pl}}~.
\eeq
For the $N_1$ to dominate the energy density of the universe prior to decays, we require that $ T_{\rm MD} >  T_{\Gamma}$. This places a requirement on the ratio of the fraction of energy in the radiation $r$ compared to the fraction of energy in $N_1$ (as given by $1-r$), specifically one requires that
\beq
\frac{r}{1-r}<\left(\frac{256\pi^{5}g_{\star}}{45M_{\rm Pl}^{2}}\right)^{1/3}y_{1}^{-4/3}M_{N_{1}}^{2/3} \approx 5\times 10^{11} \left(\frac{y_{1}}{10^{-12}}\right)^{-4/3}\left(\frac{M_{N_{1}}}{10^{10}}\right)^{2/3}.
\eeq
Thus unless the fraction of energy density in $N_1$ is diminutive compared to that in radiation at  $T=T_\star$, then an early period of matter domination will occur.

\end{document}